\documentclass[12pt]{spieman}  
\usepackage{amsmath,amsfonts,amssymb}
\usepackage{graphicx}
\usepackage{setspace}
\usepackage{tocloft}
\usepackage{soul,color} 
\usepackage{lineno}

\title{Ultra-miniature dual-wavelength spatial frequency domain imaging for micro-endoscopy}

\author[a]{Jane Crowley}
\author[a,*]{George S.D. Gordon}
\affil[a]{University of Nottingham, Department of Electrical and Electronic Engineering, Optics and Photonics Group, University Park, Nottingham, United Kingdom, NG72RD}

\cftpagenumbersoff{figure}
\cftpagenumbersoff{table} 
\begin{document} 
\maketitle

\begin{abstract}

There is a need for a cost-effective, quantitative imaging tool that can be deployed endoscopically to better detect early stage gastrointestinal cancers. Spatial frequency domain imaging (SFDI) is a low-cost imaging technique that produces near-real time, quantitative maps of absorption and reduced scattering coefficients, but most implementations are bulky and suitable only for use outside the body. We present an ultra-miniature SFDI system comprised of an optical fiber array (diameter $0.125$ mm) and a micro camera ($1\times1$ mm package) displacing conventionally bulky components, in particular the projector. The prototype has outer diameter $3$ mm, but the individual components dimensions could permit future packaging to $<1.5$ mm diameter. We develop a phase-tracking algorithm to rapidly extract images with fringe projections at $3$ equispaced phase shifts in order to perform SFDI demodulation. To validate performance, we first demonstrate comparable recovery of quantitative optical properties between our ultra-miniature system and a conventional bench-top SFDI system with agreement of $15$\% and $6$\% for absorption and reduced scattering respectively. Next, we demonstrate imaging of absorption and reduced scattering of tissue-mimicking phantoms providing enhanced contrast between simulated tissue types (healthy and tumour), done simultaneously at wavelengths of $515$ nm and $660$ nm. This device shows promise as a cost-effective, quantitative imaging tool to detect variations in optical absorption and scattering as indicators of cancer. 
\end{abstract}


{\noindent \footnotesize\textbf{*}\linkable{George.Gordon@nottingham.ac.uk} }

\begin{spacing}{2}   

\section{Introduction}
\label{sect:intro}  
Gastrointestinal cancers account for one quarter of the global cancer incidence and over one third of all cancer related deaths \cite{Arnold2020}. Wider population-based endoscopic screening can significantly decrease mortality \cite{Wei2015}, but miss rates for some types of polyps during diagnostic colonoscopies can be as high as $26$\% \cite{Zhao2019}. There is therefore a need for an improved contrast imaging device that is minimally invasive and can be deployed endoscopically.  To be deployed via the instrument channel of a standard colonoscope it must have an outer diameter $<3$ mm, and to be suitable for population screening programmes it must be relatively low-cost to manufacture and operate.

Spatial frequency domain imaging (SFDI) is a low-cost imaging technique that returns quantitative maps of absorption and reduced scattering coefficients in close to real time \cite{Cuccia2005, Cuccia2009}. SFDI requires a $2$D illumination pattern of known spatial frequency to be generated and projected onto a sample of interest, with the result captured on a standard CMOS camera. Demodulation is then performed to obtain the high and low frequency modulation amplitudes by capturing three separate patterns equally shifted in phase ($I_1, I_2$,and $I_3$) using the equations:
\begin{equation}
    I_{AC}(x_i) = \frac{\sqrt{2}}{3}[(I_1(x_i) - I_2(x_i))^2 + (I_2(x_i) - I_3(x_i))^2 + (I_3(x_i) - I_1(x_i))^2]^{1/2}
    \label{Eqn:MAC}
\end{equation}
\begin{equation}
    I_{DC}(x_i) = \frac{1}{3}[I_1(x_i) + I_2(x_i) + I_3(x_i)]
    \label{eqn:MDC}
\end{equation}
This is repeated on a reference material of known optical properties (and hence known diffuse reflectance values) such that the modulation transfer function of the imaging system can be corrected for, and diffuse reflectance values obtained. Diffuse reflectance values are then used to estimate absorption and reduced scattering coefficients using a look-up table generated from either the Diffusion Approximation or Monte Carlo simulation as solutions to the radiative transfer equation. Obtaining the optical properties at more than one wavelength allows for the extraction of additional tissue information, such as chromophore concentration via the Beer-Lambert law. This addition of endogenous contrast information is an aid in diagnosing tissue types, and has been used during breast reconstructive surgery for oxygenation imaging \cite{FirstInHumanGioux}.  

Performing SFDI at more than one wavelength simultaneously is advantageous for several reasons. Firstly, it gives the opportunity to penetrate to different depths in the sample of interest with different wavelengths\cite{Saager2011}. Secondly, it introduces the capability to obtain chromophore information, such as oxyhaemoglobin and deoxyhaemoglobin concentration, by measuring the variation in absorption coefficient at more than one wavelength\cite{Yao2013}\cite{Mazhar2010}. Blood oxygenation SFDI systems often operate in the Red/IR e.g. \cite{Chen2020}, but most fiber bundle systems operate in the green to avoid too much cross-coupling between fibers \cite{Chen2008_coretocore}. Also, using just one individual wavelength (e.g. $515$ nm), one can obtain different structural tissue information.

To improve speed of SFDI systems towards real-time operation, a single phase image can be used instead of three: a technique termed single snapshot of optical properties (SSOP)\cite{Vervandier2013}. SSOP uses a Fourier demodulation method to perform spatial frequency demodulation which typically results in poorer image quality, although a method to retain more spatial resolution has been proposed using a Hilbert Transform instead \cite{Nadeau2014}, and
emerging convolutional neural network techniques can improve resolution \cite{Chen2021, Aguenounon2020}. SSOP has been shown to successfully quantify bowel ischaemia\cite{RodriguezLuna2022}. 

SFDI has shown to return successful contrast between healthy and malignant resected esophageal and colon tissue \cite{Sweer2019, Nandy2018}. \emph{Sweer et al.} imaged resected esophageal tissue from eight patients undergoing esophagectomy. By comparing regions imaged with a commercially available SFDI system from \emph{Modulim} \cite{SFDImodulim} with results from histological analysis of tissue, it was determined that healthy esophageal tissue has a reduced scattering coefficient higher than the reduced scattering coefficient of both invasive squamous cell carcinoma and Barrett's esophagus with mild chronic inflammation, in the red wavelength region $\sim 660$ nm. The absorption coefficient of healthy esophageal tissue is lower than invasive squamous cell carcinoma and analogous to that of Barrett's esophagus with mild chronic inflammation, at a wavelength of $\sim660$ nm. \emph{Nandy et al.} found that healthy colon tissue has a higher reduced scattering coefficient than malignant colon tissue and a lower absorption coefficient across the wavelength range $460-630$ nm.

SFDI is an attractive choice for an imaging modality because it does not require high-powered lasers, sensitive detectors (mobile phone cameras are sufficient) or complex optical components.  It is therefore relatively low-cost to manufacture and operate devices, and they can be miniaturized easily.  As a result, a number of SFDI systems exist, such as large commercial systems\cite{SFDImodulim}, portable handheld systems\cite{Saager2017}, handheld $3$D printed systems\cite{Erfanzadeh2018}, compact multispectral imaging systems\cite{Kennedy2021}.  However, in most existing systems the projector element remains costly and difficult to miniaturize, being typically comprised of either a digital micromirror device (DMD) projector \cite{Saager2017}, a motorized grating \cite{Erfanzadeh2018, Kennedy2021}, or a static spatial frequency transparency \cite{VandeGiessen2015}.  

There have been a number of approaches to miniaturize SFDI projectors to make them suitable for endoscopic deployment.  Fixed gratings have been used to achieve SSOP via rigid endoscopes \cite{Angelo2017}. While SSOP is advantageous as it reduces acquisition times, it poses several disadvantages, such as reduced image quality due to the use of filtering a single image. The previously developed probe is rigid in nature and not suitable for imaging in the gastrointestinal tract. Fixed gratings have also been used for optical sectioning via flexible imaging fiber bundles \cite{Ford2012}. The use of a micro camera is advantageous over imaging through a fiber bundle as an imaging fiber bundle is highly sensitive to vibrations, cross coupling and fiber movements, making the reconstruction of images challenging. Phase-shifted illumination has been demonstrated via an imaging fiber bundle \cite{Supekar2022}, but the use of DMDs is relatively high cost, and commercial fiber bundles projection only support high fidelity fringes at green wavelengths due to increased cross-coupling between cores in red wavelengths\cite{Chen2008_coretocore}. Ultra-thin fiber arrays have been used to create fringe patterns interferometrically for profilometry but not, to our knowledge, for SFDI\cite{Zhang2013Fiber-opticSystem}.  

None of these existing systems are suitable for routine endoscopic deployment in the gastrointestinal tract because they either use DMD-based projectors which are costly and cannot be sufficiently miniaturized; use fiber bundles which produce low-quality fringe patterns at a limited set of wavelengths and only record low resolution images; or use rigid endoscopes which are not flexible enough. 

We have therefore developed an ultra-miniature SFDI system, with an outer diameter $3$ mm that uses a fiber array to interferometrically produce fringe patterns at green ($515$ nm) and red ($660$ nm) wavelengths and records images at $320 \times 320$ pixel resolution using a micro camera.  The prototype packaging is sufficiently small that is it compatible with the instrument channel of a standard colonoscope. This makes the device comparable to the thinnest previous SFDI system designs that used fiber bundles to achieve a total diameter of $2.7$ mm\cite{Ford2012}.

We first compare optical property measurements in our ultra-miniature system to that of a conventional bench-top system and find agreement between absorption and reduced scattering coefficients of $15$\% and $6$\% respectively. We show the potential to operate the system at more than one wavelength simultaneously, enabling rapid tissue property measurements. This device therefore shows potential to be deployed endoscopically for \emph{in-vivo} gastrointestinal imaging to detect optical properties as potential indicators of cancer.

\section{Methods}
\label{sect:Methods}
\subsection{Component design and selection}
\label{subsect: Component selection}
The primary components needed for an SFDI system are a source of pattern projection and an image detector to capture the projected patterns on a sample of interest. We chose to use an optical fiber array as the source of projection patterns and a micro camera as the detector. 

\begin{figure}[!htbp]
    \centering
    \includegraphics[width=1\linewidth]{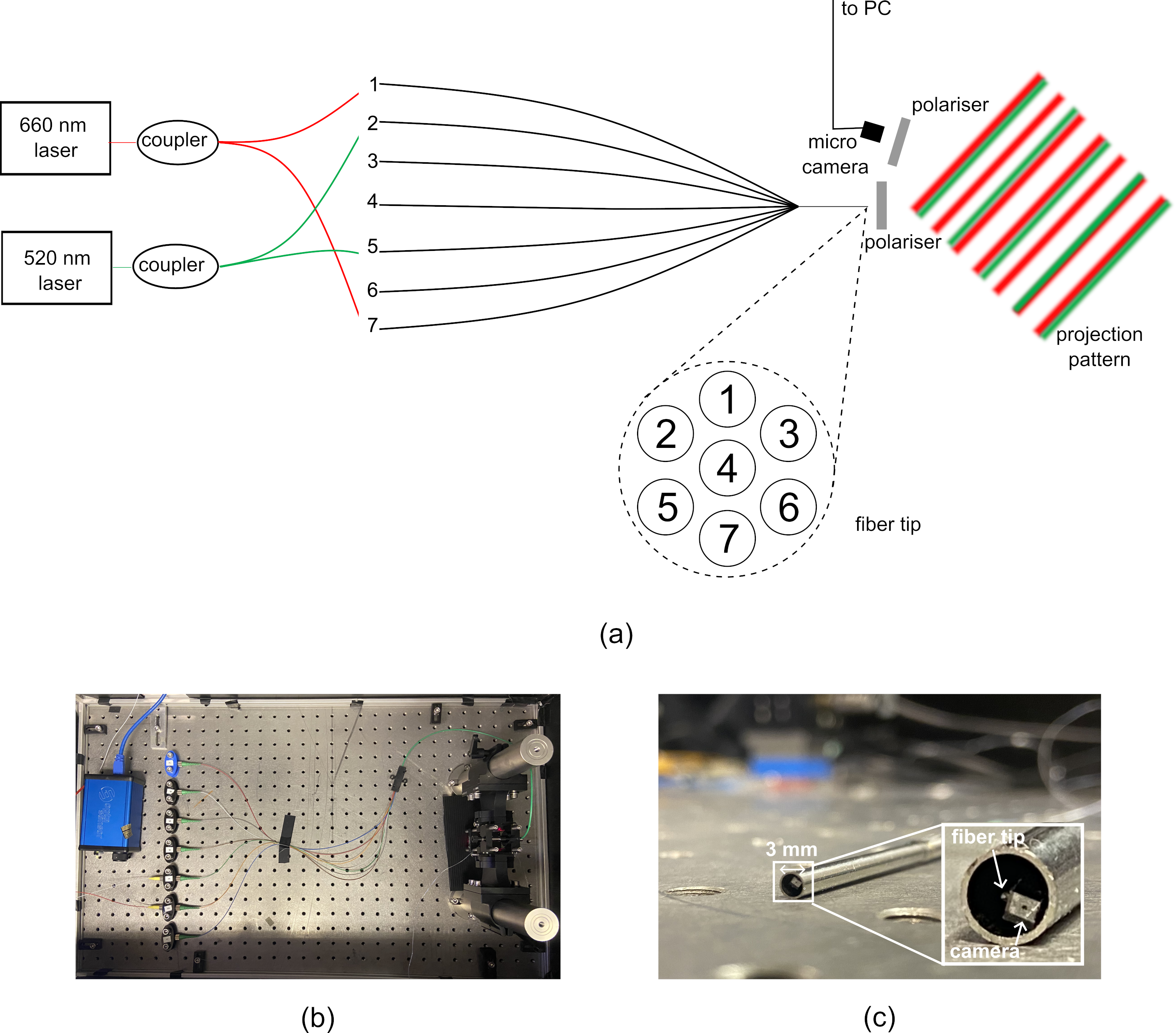}
    \caption{Proposed ultra-miniature SFDI system (a) Schematic of fiber array in ultra-miniature SFDI system showing dual wavelength illumination simultaneously. Light passes from the two lasers into the fiber array via a selection of 7 single-mode fiber input ports. At the tip of the fused taper, the fibers are spaced in a hexagonal array, providing three possible spacings. Crossed polarizers are placed in front of the fiber tip and the micro camera to reduce specular reflections from the imaging sample. (b) Photograph of experimental set up. (c) Prototype device package of $3$ mm diameter with inset showing zoomed in view of fiber tip and camera.}
    \label{fig:fibberarray_schematic}
\end{figure}

\begin{figure}[!htpb]
    \centering
    \includegraphics[width=1\linewidth]{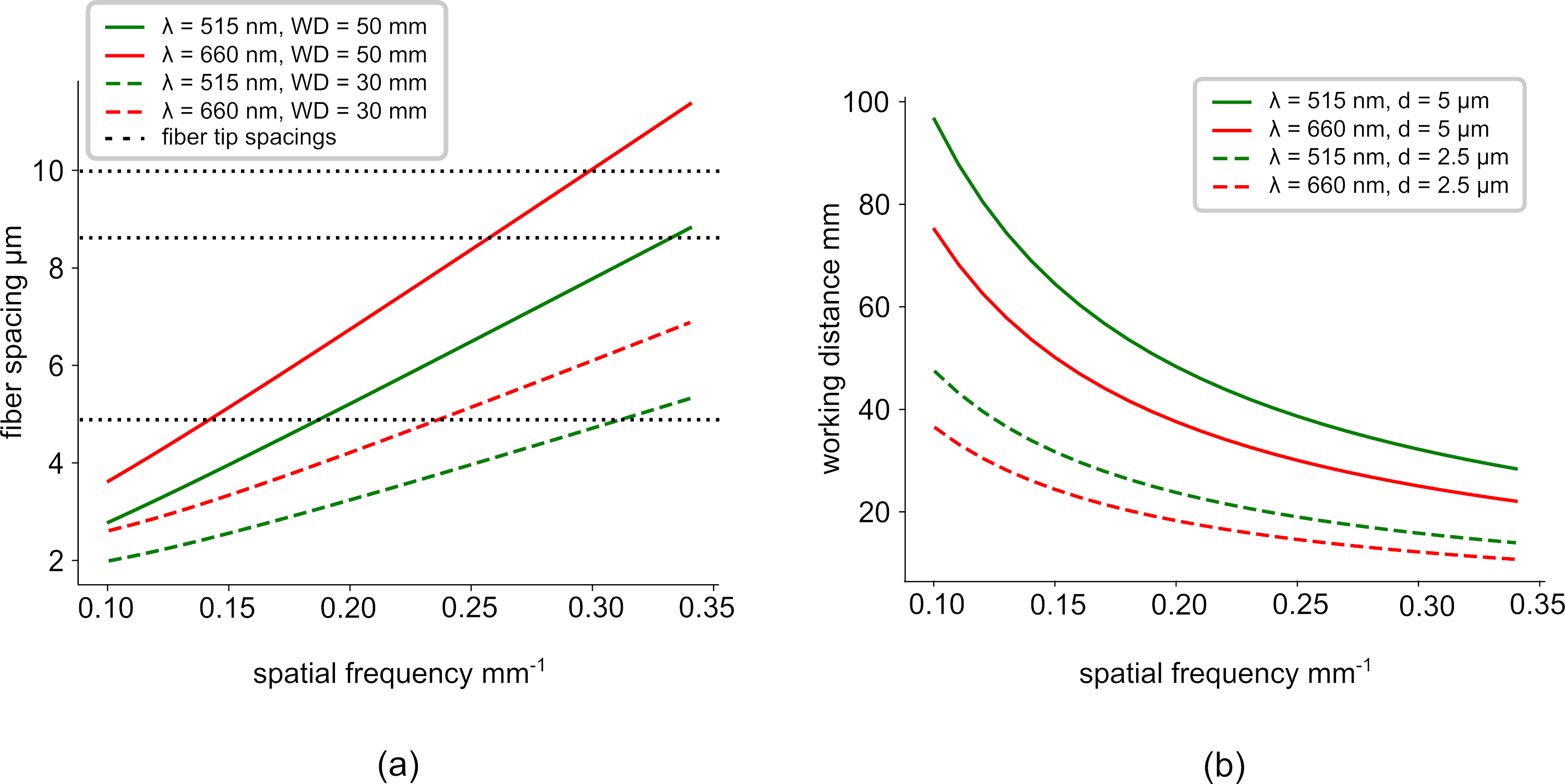}
    \caption{Determining desired fiber spacing to produce spatial frequencies within our range of interest $0.1-0.3$ mm$^{-1}$ (a) Addressable spatial frequency projection at working distances (WD) of $50$ mm (solid lines) and $30$ mm (dashed lines). The dotted lines represent the three possible fiber tip spacings of $5$, $8.66$ and $10$ $\mu$m  (b) proposed design of spatial frequency projection at various working distances for fiber tip spacing (d) of $5$ $\mu$m (solid lines) and $2.5$ $\mu$m (dashed lines), useful for smaller working distances.}
    \label{fig:FiberArray_spacingcalc}
\end{figure}

To create an ultra-miniature fringe projector without using DMD elements, we designed a customized two-dimensional pitch-reducing optical fiber array (PROFA\texttrademark, \emph{Chiral Photonics}, NJ) to create fringes interferometrically, shown in Fig. \ref{fig:fibberarray_schematic}. The fiber array was designed to produce interference patterns within a widely used spatial frequency range ($0.1-0.3$ mm$^{-1}$ \cite{Gioux2019, Hayakawa2018}) at an initial test working distance of $50$ mm when two adjacent channels are illuminated by the same laser source. To compute the required fiber spacings, we used a double slit equation:
\begin{equation}
    m\lambda = d \sin{\theta}
    \label{eqn:double slit}
\end{equation}
where $m$ is the number of the interference line spacings from the central point, $\lambda$ is the wavelength of light, $d$ is the distance between slits and $\theta$ is the angle of projection. The desired wavelength was chosen to be $660$ nm. The distance from slit to projection pattern, i.e. the working distance, was chosen initially to be $50$ mm, which is the maximum working distance of the camera. Using Eqn \ref{eqn:double slit}, we can therefore determine the spacing $d$ required to produce our spatial frequencies of interest. The fabricated fiber array has spacings of $5$, $8.66$ and $10$ $\mu$m, which will produce spatial frequencies of $0.15$, $0.25$, $0.3$ mm$^{-1}$ at 660nm, as shown in Fig. \ref{fig:FiberArray_spacingcalc} a. We can then determine the spatial frequency projection at varying fiber to sample working distances, shown in Fig \ref{fig:FiberArray_spacingcalc} (b). Typical endoscope working distances are $20-30$ mm \cite{EndoscopeFieldofView}, which is achievable using the 5$\mu$m spacing option of our array with $0.3$ mm$^{-1}$ spatial frequency, though in future designs a 2.5$\mu m$ spacing could enable even shorter working distances. The 7 fiber channels are spaced at the tip as shown in Fig \ref{fig:fibberarray_schematic}. The light sources used are a $5$ mW $660$ nm laser diode (LPS-660-FC, \emph{Thorlabs}) and a $3$ mW $515$ nm laser (LP515-SF3, \emph{Thorlabs}).

The camera chosen is a $1\times1$ mm micro camera module (Osiris M module, \emph{OptaSensor}, Germany). The camera has a resolution of $320\times320$ pixels, with an individual pixel size of $2.4$ $\mu$m. An in-built lens placed in front of the sensor provides  horizontal and diagonal field of views of $68^{\circ}$ and $90^{\circ}$ respectively, accompanied by a depth of focus of $5-50$ mm. The camera module produces a $12$ bit RGB raw image output. The camera is accompanied by software to control camera parameters such as exposure, gamma correction and black level correction. The automatic exposure correction was disabled so that all image frames contain the same optical power ranges. The micro camera has a frame rate of up to $50$ fps, but here we typically operate it at $10$ fps due to capture software limitations.  However, $10$ fps is the minimum rate required for proper endoscopic visualization\cite{Venugopal2013}.  

To minimize specular reflections present on the imaging sample, adhesive-backed polymer polarizer sheets and placed in front of the camera and fiber tip to create cross-polarization. The camera is also placed at a small angle of $4^{\circ}$ to the fiber to further limit specular reflections on the imaging sample. This angle is smaller than conventional SFDI systems\cite{FirstInHumanGioux}, but is more amenable to miniaturization.  Previous work has shown that this angle can still produce high quality optical property maps\cite{Crowley2023Blender}

\subsection{Phase-tracking algorithm}
\label{subsect: Characterisation of fringes}
An inherent property of an interferometer such as our fiber array is that the sinusoidal pattern produced will shift over time due to mechanical drifts, vibrations, temperature and intensity variations \cite{Kinnstaetter1988}. Conventional wisdom may suggest using a complex set up consisting of a phase-shifting control system and a piezoelectric transducer driver to stabilize and control this phase shift\cite{Zhang2013Fiber-opticSystem}.  However, we exploit the natural phase drift to our advantage via a phase-tracking algorithm.

\begin{figure}[!htpb]
    \centering
    \includegraphics[width=1\linewidth]{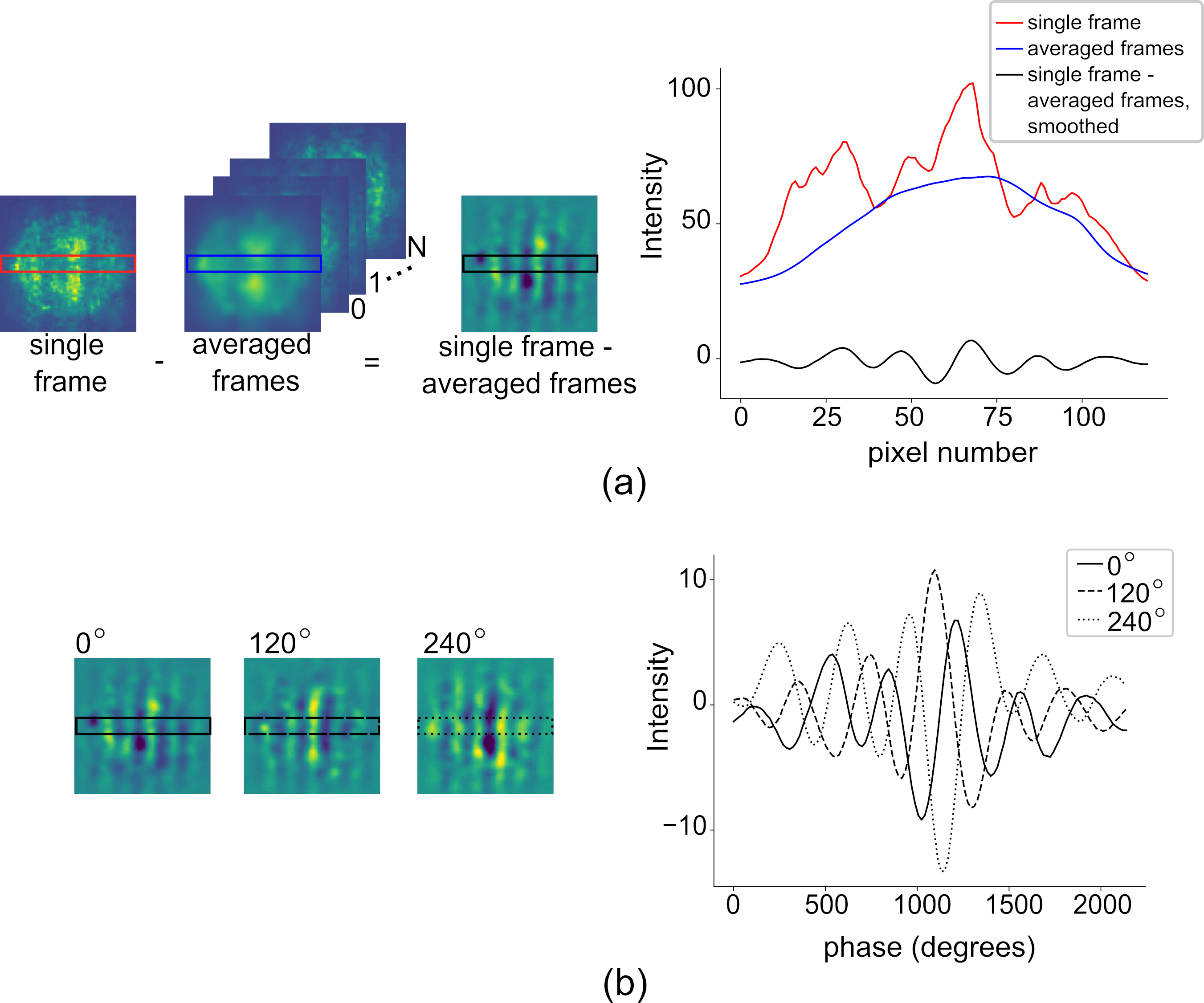}
    \caption{Characterization of fringes and phase tracking (a) image of selected zeroth frame and average of all frames, where $N$ is the total number of all frames within video capture, and corresponding cross sections. The single frame - averaged frames has had a smoothing filter applied (b) image of zeroth frame, $120^{\circ}$ shifted frame and $240^{\circ}$ shifted frame and corresponding cross sections, all with smoothing filters applied.}
    \label{fig:fringe_characterisation}
\end{figure}

A video, typically $10-20$ s, is first recorded of the shifting sinusoidal pattern on a sample of interest. This timeframe is purely selected for experimental convenience but in practice only the first second or so of recorded frames is required. To determine which frames to use for demodulation, we first take an average of all frames within the video and subtract this from each individual frame. This allows us to visualize the spatial frequency pattern with reduced noise (see Fig \ref{fig:fringe_characterisation} a).

We then take an average across several rows within the frame, applying a smoothing filter, and plot the sinusoidal pattern. We select a zeroth frame and designate the phase of the extracted sinusoid to be $0^{\circ}$.  Next, we calculate the average distance between adjacent maxima of this sinusoid in pixels.  This value gives the period of the pattern, in pixel units. Custom \emph{Python} code then cycles through all frames in the captured video and selects frames of equal intensity variation whose sinusoidal projections have relative phases of $(120\pm 10)^{\circ}$ and $(240\pm 10)^{\circ}$ from the selected zeroth frame (see Fig \ref{fig:fringe_characterisation} b). Frames where the sine wave is non-discernible or the intensity variation between peak and trough is low relative to the zeroth frame are disregarded. This eliminates frames where coherence is temporarily disturbed while perturbations are still in progress. We then select these frame numbers from the initial video and demodulate the images using Eqns \ref{Eqn:MAC} and \ref{eqn:MDC}.

\subsection{Imaging homogeneous tissue-mimicking phantoms}
In order to perform initial validation of the system, we fabricated tissue mimicking co-polymer in oil phantoms with tunable optical properties by controlling concentrations of TiO$_2$ and Nigrosin dye \cite{Hacker2021}. The fabricated phantoms had a thickness of $30$ mm and were ensured to be non-transparent so as to meet the semi-infinite thickness requirement of SFDI \cite{Liemert2013}. We fabricated two phantom batches; one with increasing amounts of dye stock solution from $0.5 - 1$ g ($0.0015-0.0030$ w/v\%) corresponding to an absorption coefficient range of $0.006 - 0.017$ $mm^{-1}$ at $660$ nm and the second with increasing amounts of TiO$_2$ from $0.07 - 0.13$ g ($0.07 - 0.13$ w/v\%) corresponding to a reduced scattering coefficient range of $0.52 - 0.99$ $mm^{-1}$ at $660$ nm. The batch with increasing dye stock solution each had $0.1$ g of TiO$_2$ and the batch with increasing TiO$_2$ each had $0.5$ g of dye stock solution to ensure the semi-infinite material requirement was met. We chose these optical property ranges as they lay within optical properties of interest of typical gastrointestinal tissue samples \cite{Sweer2019} and they had previously been calibrated for in literature using a double integrating sphere (DIS) \cite{Hacker2021}. 

For purposes of comparison, we imaged phantoms in both our ultra-miniature system and a standard bench-top SFDI system. The bench-top system was built in-house using a commercial projector (\emph{LG Minibeam PH150g HD ready mini projector}), a Raspberry Pi camera, crossed polarizers placed in front of the camera and projector, and a $635$ nm filter placed in front of the camera to ensure only red light was captured. The system was validated against phantoms with optical properties measured in a double integrating sphere system and found to be in agreement with a margin of error around 20\%, which is within the typical expected range \cite{DongHu2018}.

The phantoms were placed in the bench-top system such that the projector is placed at a $6^{\circ}$ angle to the plane of the camera, the projector-camera distance is fixed at $35$ mm, and the distance from the projector-camera plane to the base breadboard is $220$ mm. The Diffusion Approximation was used to generate a look-up table to recover the optical properties of the phantoms. Because our ultra-miniature system uses a laser at $660$ nm and the bench-top system uses white-light source with a camera filter at $635$ nm, a small adjustment for wavelength is required.  First, separate LUTs are computed for each of the wavelengths.  Second, an adjustment based on multiwavelength measurements from a double integrating sphere system is applied \cite{Hacker2021}. For these particular phantoms, the measured difference via DIS between absorption and reduced scattering coefficients from $635$ nm to $660$ nm is $14$\% and $3$\% respectively. Therefore, we adjust reference optical properties for optical property calculation depending whether we were using the bench-top or ultra-miniature system. When imaging phantoms at $515$ nm in the ultra-miniature system, the LUT and reference optical properties were also adjusted for accordingly. 

We placed the phantoms such that the top of the phantom was $50$ mm from the distal end of the imaging probe and the projection pattern was in the center of the sample. We took videos of the shifting projection pattern on the phantom for $10-20$ s. The video was then input to \emph{Python} phase-tracking code for processing described in Sect \ref{subsect: Characterisation of fringes} to find the exact frames needed to calculate the optical properties. We imaged each phantom at three different spatial frequencies by illuminating three different fiber channel combinations. We calculated the optical properties using a look-up table generated from the Diffusion Approximation. For each phantom, we calculated the optical property maps a total of $18$ times, using every other phantom as a reference in turn for each spatial frequency. This approach helps to average out errors arising from mismatches in expected optical properties of phantoms, which arises in turn due to discrepancies between DIS and SFDI measurements, which can be up to $20$\% \cite{DongHu2018}. Finally, the mean of all $18$ optical property maps is used to determine the absorption and reduced scattering coefficients. A 2D Gaussian filter with standard deviation of $5$ pixels was applied to resultant optical property maps using \emph{scipy.ndimage.gaussian\_filter}.

\subsection{Dual wavelength imaging}
Multi-wavelength imaging is possible with this system as the fiber array consists of seven channels and only two are needed per wavelength to produce an interference pattern. Therefore, this system has the potential to explore up to tri-wavelength simultaneous illumination. We imaged three phantoms with $660$ nm projection only, then $515$ nm projection only, and finally with $660$ nm and $515$ nm projected simultaneously. We perform dual-wavelength imaging by illuminating channels $1\&7$ with $660$ nm and channels $2\&5$ with $515$ nm, producing spatial frequency patterns of $0.3$ mm$^{-1}$ and $0.2$ mm$^{-1}$ respectively at a $50$ mm working distance. A video is captured of both illumination patterns simultaneously, and analysis is carried out by extracting the red and green channels from the video capture. Following the same process in Sect \ref{subsect: Characterisation of fringes}, fringes of interest are selected and optical properties calculated. Expanding the existing system to tri-colour would be possible either by adding an additional laser of, say $\sim450$ nm, to two available illumination channels will produce a spatial frequency of $0.22$ mm$^{-1}$, and could be analysed from the blue channel of the captured video. Multi-wavelength imaging would probe different depths and could be used to image optical properties of layered material. 

\subsection{Extrapolating performance to clinical applications}
The ultra-small size of our prototype device may incur performance penalties compared to standard bench-top SFDI.  Qualitatively, this appears as noise in measured images.  However, the ultimate goal of the tool is to provide sufficient contrast for identifying diseased tissue compared to healthy tissue. To assess this, we post-process our optical property measurements to estimate the sensitivity and specificity that could be achieved in a clinical device. This is then compared against American Society of Gastrointestinal Endoscopy (ASGE) guidelines, which require a new device to demonstrate $90$\% sensitivity and $80$\% specificity in order to be used in lieu of biopsy sampling \cite{ASGE2012}.

To extrapolate our device's potential clinical performance we simulate random measurements based on previously measured absorption and scattering statistics of healthy esophageal tissue, Barrett's esophagus and squamous cell carcinoma (SCC) \cite{Sweer2019}. Because of the non-negativity and underlying multiplicative nature of absorption and scattering processes, a log-normal distribution is used to generate samples\cite{limpertLognormalDistributionsSciences2001}. A baseline model is first created that estimates performance under the assumption that intra-sample variation is purely biological in origin. To adapt this into the model used here, we add an additional term to the variance based on error measurements observed from our ultra-miniature system.

500 random training samples are generated for the healthy, Barrett's and SCC tissue types, and two binary linear support vector machine classifiers are trained (healthy vs. Barrett's and healthy vs SCC). A further 500 validation samples are then generated for each of the 3 classes and passed through the trained classifiers.  Based on the classification error rates, sensitivity and specificity for the baseline and ultra-miniature system are estimated.

\section{Results}
\label{sect:Results}
\subsection{Projector performance}
The expected spatial frequency of the projected illumination pattern is comparable to the desired spatial frequency with $12$\% and $7$\% error for $660$ nm and $515$ nm respectively. Some channels produce clearer interference patterns than others, due to cross talk between fibers. This also results in the interference pattern from some channels being more stable than others in time, with interference patterns being stable for $<1$ s under typical operating conditions, but $>10$ s if the fibers are kept still.

\begin{figure}[!htpb]
    \centering
    \includegraphics[width=1\linewidth]{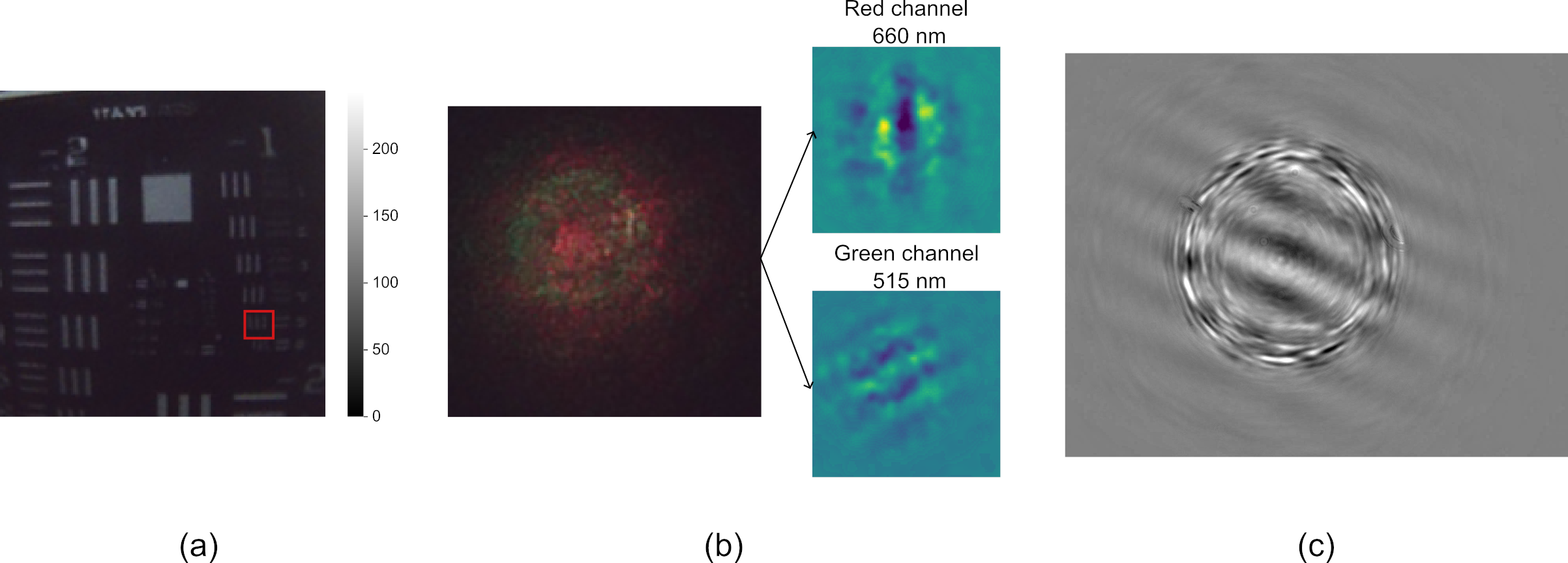}
    \caption{Raw performance of camera and projector (a) Image of USAF target captured with mini camera module taken with room lights on (b) image captured with mini camera module of dual wavelength projection from fiber tip showing extracted red and green channels respectively, (c) Direct capture on image sensor of projected fringe pattern, showing good fringe contrast within a region of circular interference.}
    \label{fig:cam_fiber_results}
\end{figure}

Through imaging a resolution target (R3L3S1N - Negative 1951 USAF Test Target, 3'' x 3'', \emph{Thorlabs}, UK), we determined the resolution of the imaging system to be $0.793$ lp/mm at a working distance of $50$ mm, shown in Fig \ref{fig:cam_fiber_results} a. Fig \ref{fig:cam_fiber_results} b shows the raw performance of the projection fiber with green and red wavelengths onto an absorbing and scattering phantom. This represents a typical raw image required from the camera.  Finally, Figure \ref{fig:cam_fiber_results}c shows a direct capture on a CMOS image sensor of the projected pattern from the fiber. This shows that clear fringes are produced within an envelope, but that there is also some noise arising from aperture and cross-talk effects forming a ring around the target area.  We therefore restrict our analysis to regions within this ring where fringes are projected with high quality.

\subsection{Phase-tracking stability}
\begin{figure}[!htpb]
    \centering
    \includegraphics[width=1\linewidth]{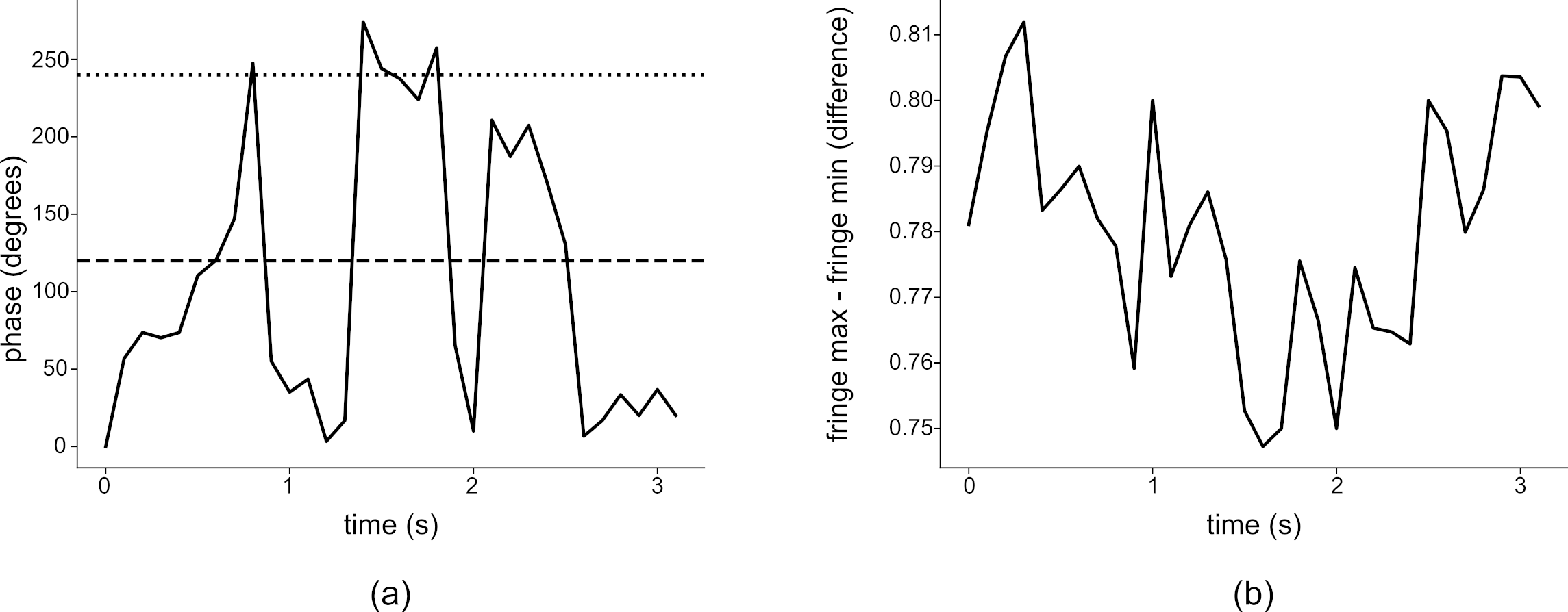}
    \caption{Fiber array performance (a) phase of frames vs time (b) fringe contrast, defined as difference from maxima to minima of interference fringes detected vs time}
    \label{fig:phase_and_fringecontrast_vs_time}
\end{figure}

We analysed the fringe phase shift and contrast for 15 recorded videos, during which we continually perturbed the fiber by hand to simulate realistic usage.  An example trace of fringe phase vs. time is shown in Fig \ref{fig:phase_and_fringecontrast_vs_time} (a) which indicates that the required 3 phases can be obtained under 1s. Fig \ref{fig:phase_and_fringecontrast_vs_time} (b) shows the calculated difference from maximum to minimum of the interference fringes as a function of time, i.e. the fringe contrast. The contrast appears relatively stable over the $3$ s time interval. To further analyse the combined effect, we used Poisson statistics across these 15 captures to estimate the expected time at maximum frame rate required to obtain 3 usable frames (defined as having sufficiently high contrast across 3 phases) with 99\% probability.  We found this varied from 0.42 s to 2.34 s with a mean of 1.17 s. This may be sufficient for practical imaging provided the probe can be held within the same field of view for 1-2 s.

To examine the stability of the fringe envelope we averaged all frames together for individual video captures and plotted the a cross-sectional envelope profile.  By repeating this across 5 separate video captures using a range of different phantoms, we observed how the fringe envelope varies.  Given that the envelope shape is largely determined by the fibre exit aperture we would expect the shape to be relatively uniform across different samples, although effects such as cross-talk between fibres may cause a variation.  Additionally, fluctuations in laser power, laser polarization state or small transients in relative power in different fibres due to manual perturbations may influence the envelope.

\begin{figure}[!htpb]
    \centering
    \includegraphics[width=0.7\linewidth]{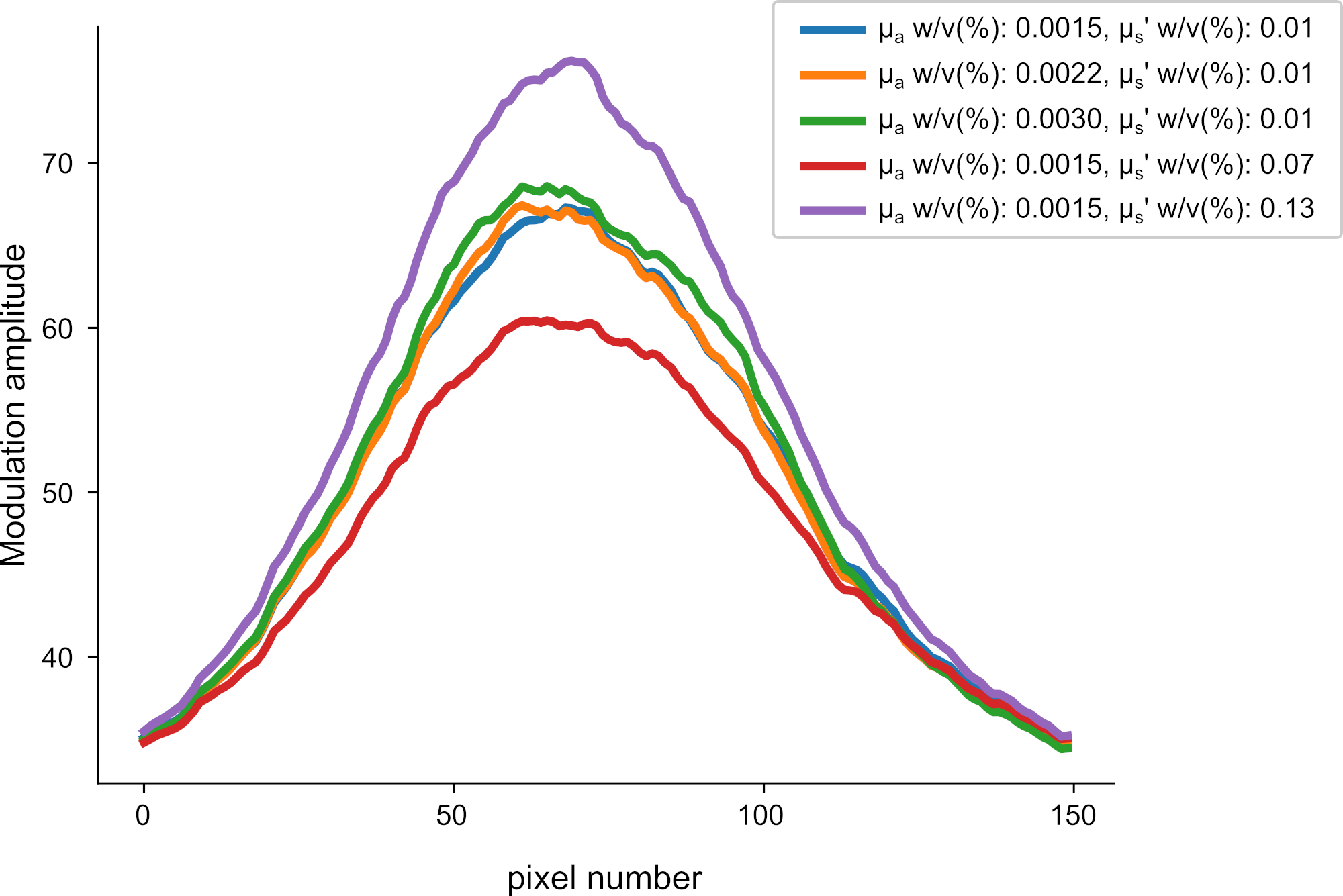}
    \caption{Plot of average fringe envelope for 5 different captures using phantoms with different optical properties.  Though the amplitude of the envelop varies in intensity, the position and shape stay relatively constant.}
    \label{fig:envelope_char}
\end{figure}

From our results, shown in \ref{fig:envelope_char}, we observe that the shape of the envelope is fairly consistent across different captures, though the absolute scale does appear to change within a range of $\pm$10\%. In future systems, this could be partially compensated by having a power meter recording laser output which can be used later used for normalization.

We therefore expect that there may be some variation between the reference phantom envelope and the measurement: the envelope will be approximately correct but the amplitudes may vary which could be a contributing source to the overall system error, in the region of 10\%.

\subsection{Comparing ultra-miniature SFDI system to bench-top SFDI system}
\label{subsect:results - comparing mini system to bench top system}
\begin{figure}[!htpb]
    \centering
    \includegraphics[width=1\linewidth]{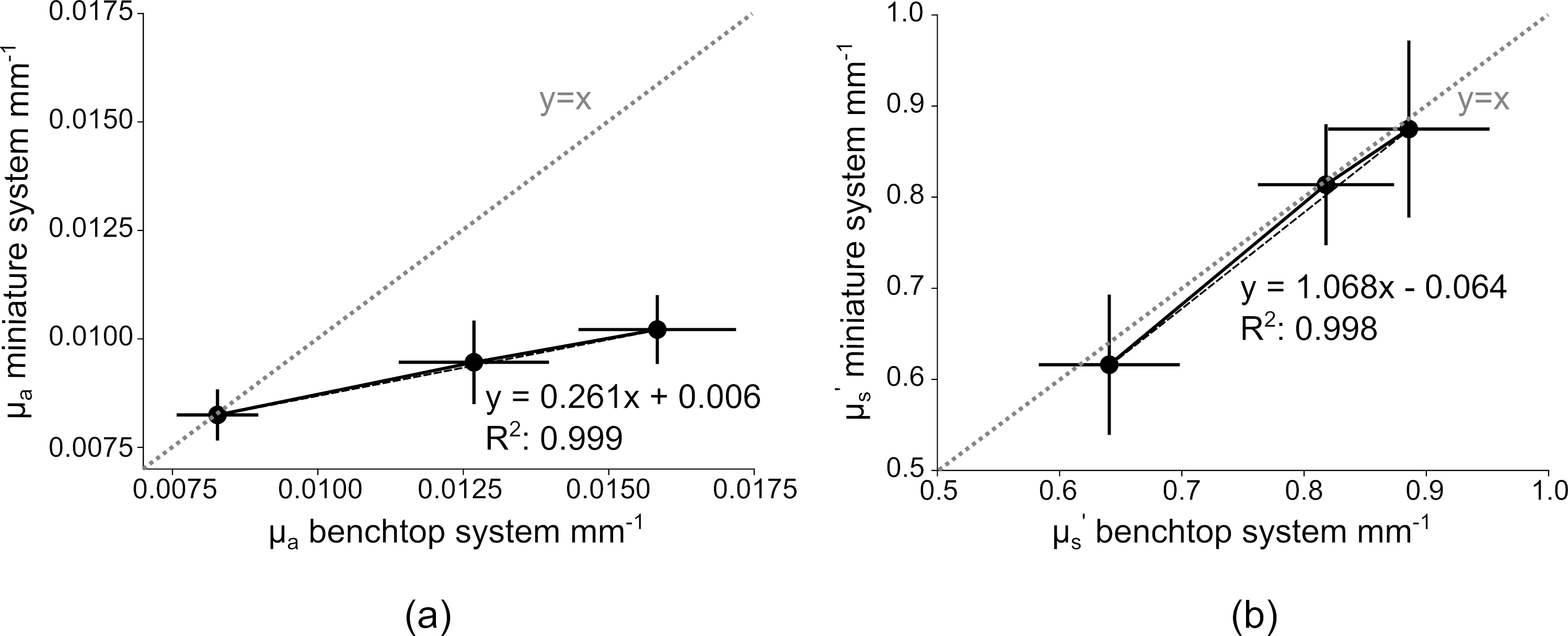}
    \caption{Comparison of bench-top SFDI system and ultra-miniature system: (a) absorption coefficient and (b) reduced scattering coefficient measured from bench-top system ($x$ axis) and ultra-miniature system ($y$ axis). Error bars represent the standard deviation across the image.}
    \label{fig:benchtop_vs_mini}
\end{figure}

We then compared optical property measurements from our conventional bench-top SFDI system to our ultra-miniature SFDI system. The results are shown in Fig \ref{fig:benchtop_vs_mini} a \& b. We found that the average standard error in absorption and reduced scattering coefficients between the ultra-miniature system and the bench-top system were $15$\% and $6$\% respectively. However, we note that the slope of the absorption graph is much less than unity, reaching as high as 60\% in the high absorption range. This may be due to a range of factors including stray background light, non-ideal sinusoid shapes and the resulting mismatch in the LUT, or varying envelope amplitudes.  Because the absorption is consistently underestimated, this may be rectified using additional calibration steps, for example phantom-based LUT generation \cite{Erickson2010} or realistic simulation of the fibre imaging set up \cite{Crowley2023Blender}.

\subsection{Imaging typical gastrointestinal conditions with ultra-miniature SFDI system}
\begin{figure}[!htpb]
    \centering
    \includegraphics[width=1\linewidth]{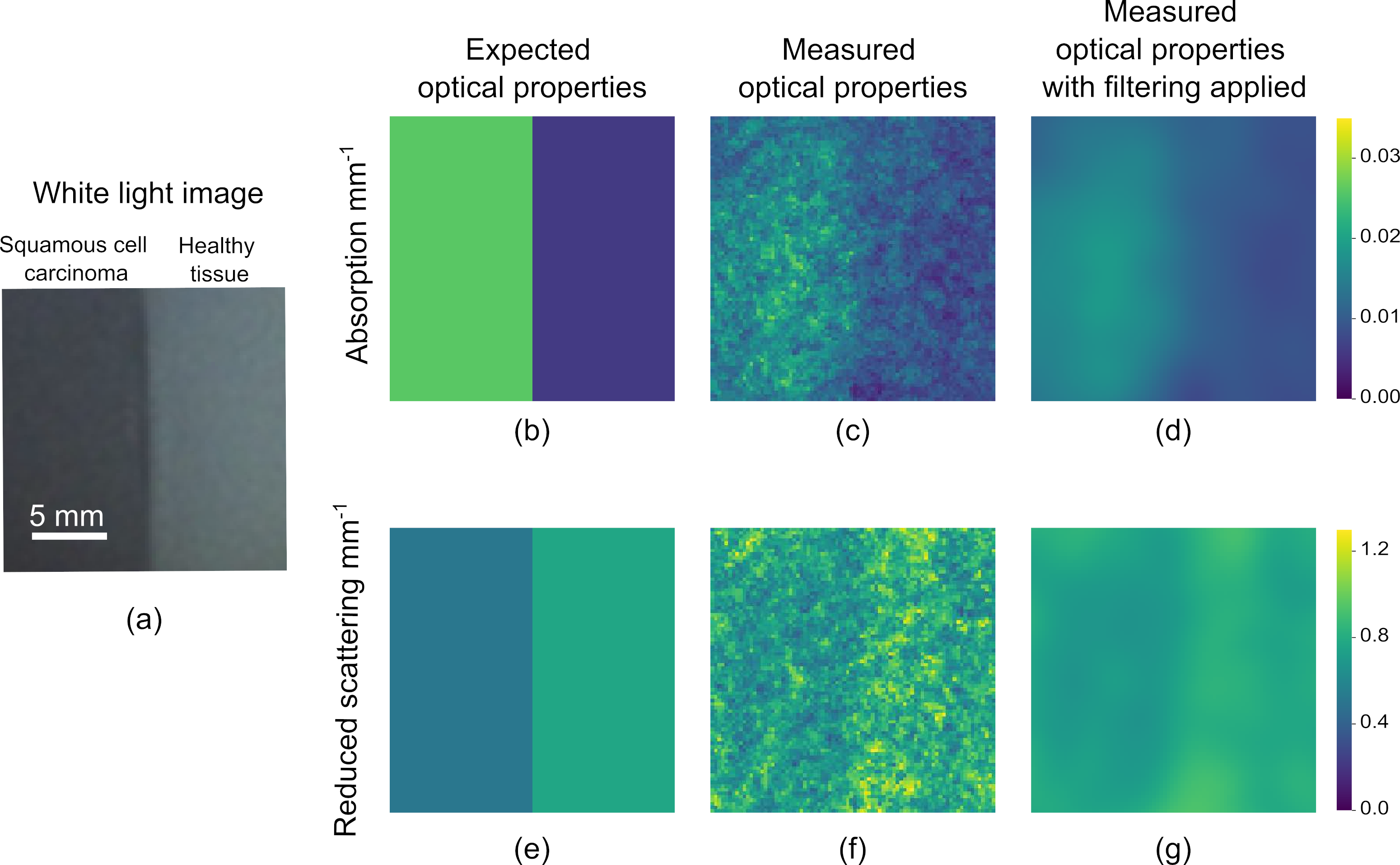}
    \caption{Imaging a phantom simulating esophageal tissue at 660nm: (a) white light image of two phantoms with different optical properties side by side (b) expected and (c) measured absorption coefficient of phantoms (d) measured absorption coefficient with smoothing filter applied (e) expected and (f) measured reduced scattering coefficient of phantoms (g) measured reduced scattering coefficient with smoothing filter applied. Expected optical properties are computed from the mean measured values of the individual phantoms measured in bench-top SFDI system.}
    \label{fig:Phantom_sidebyside}
\end{figure}

We fabricated two phantoms to simulate gastrointestinal tissue states: one with optical properties mimicking squamous cell carcinoma, and the second with optical properties mimicking healthy esophageal tissue, and placed them side by side.  SFDI imaging of this sample was then performed at 660nm, with the resulting optical property maps shown in Fig \ref{fig:Phantom_sidebyside} c \& f, and resultant optical property maps with filtering applied shown in Fig \ref{fig:Phantom_sidebyside} d \& g. 

\begin{figure}[!htpb]
    \centering
    \includegraphics[width=0.8\linewidth]{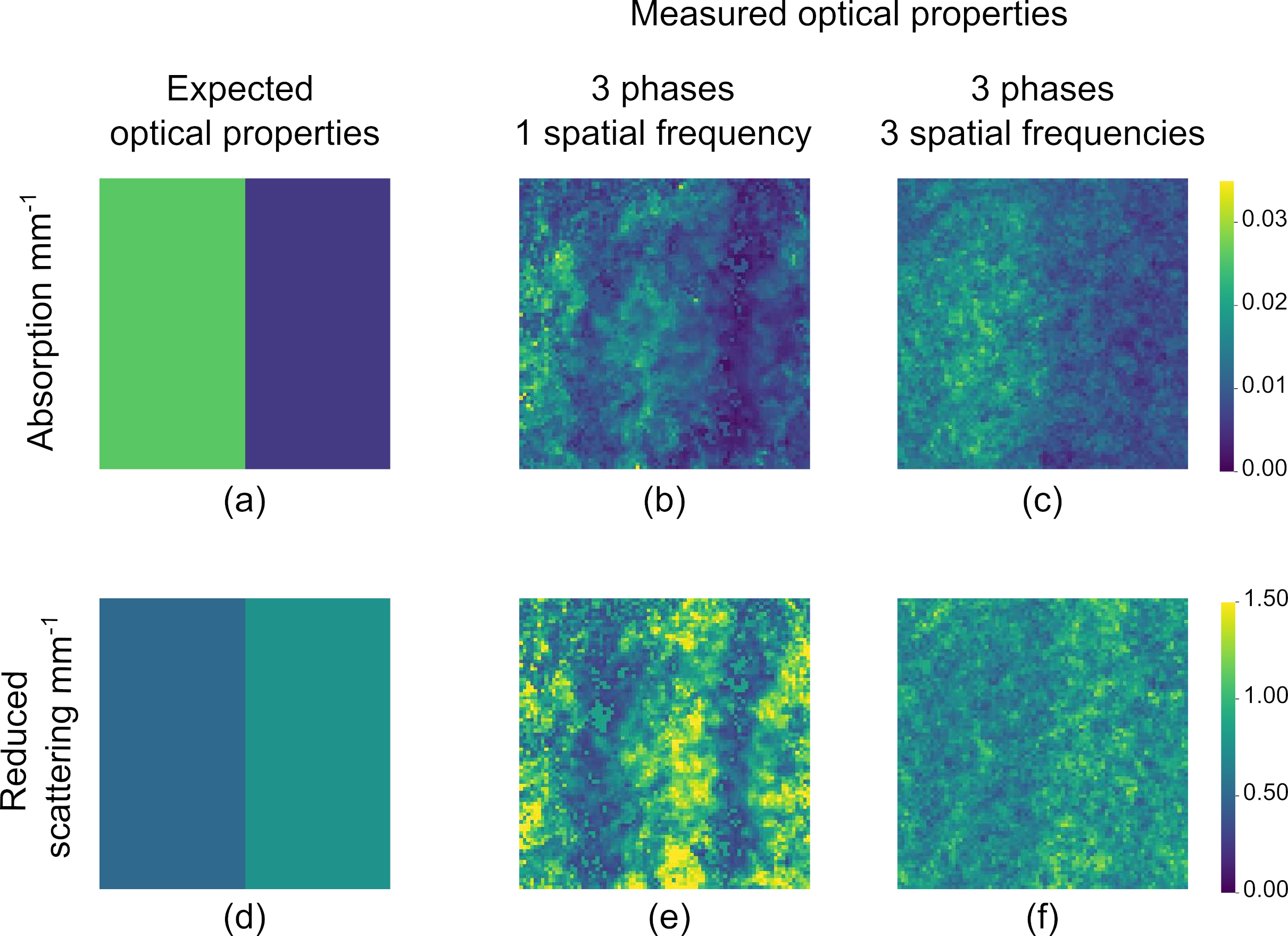}
    \caption{Visual representation of averaging over several spatial frequencies to reduce speckle noise.}
    \label{fig:speckle_fig}
\end{figure}

To illustrate the effect of speckle averaging, we compared the optical properties of the same phantoms measured with just a single spatial frequency to three spatial frequencies averaged together. The results are shown in Fig \ref{fig:speckle_fig}: Fig \ref{fig:speckle_fig}(a \& d) represent the expected optical properties of the phantoms, Fig \ref{fig:speckle_fig} (b \& e) show the optical properties measured from a single spatial frequency projection and Fig \ref{fig:speckle_fig} (c \& f) show the optical properties measured by taking the average over three different spatial frequency projections. The standard deviation of pixels in the absorption coefficient maps reduces from $0.0049$ to $0.0044$ from Fig \ref{fig:speckle_fig} (b) to (c). The standard deviation of pixels in the reduced scattering coefficient maps reduces from $0.32$ to $0.15$ from Fig \ref{fig:speckle_fig} (e) to (f).

Using our post-processing simulation, we found that for detecting Barrett's esophagus vs. healthy we could achieve a sensitivity of 76.2\% and a specificity of 82.8\%.  This falls just short of the ASGE guidelines \cite{ASGE2012}. However, for SCC vs. healthy we find $>$99\% sensitivity and 89.6\% specificity.  This meets the ASGE guidelines and is in agreement with experimental pre-clinical SFDI studies that typically achieve values $>$90\% for both sensitivity and specificity when comparing healthy and cancer tissue \cite{nandyCharacterizingOpticalProperties2016, shanthakumarComparisonSpectroscopyImaging2023}.

\subsection{Dual-wavelength imaging}
Finally, we characterized the performance across the two wavelengths.  We found that the recovered optical properties varied by $\leq 10\%$ when the two wavelengths are measured simultaneously, compared to measuring them sequentially. This demonstrates the capability of the system to image optical properties at two wavelengths simultaneously with relatively low cross-coupling. 

\begin{figure}[!htpb]
    \centering
    \includegraphics[width=1\linewidth]{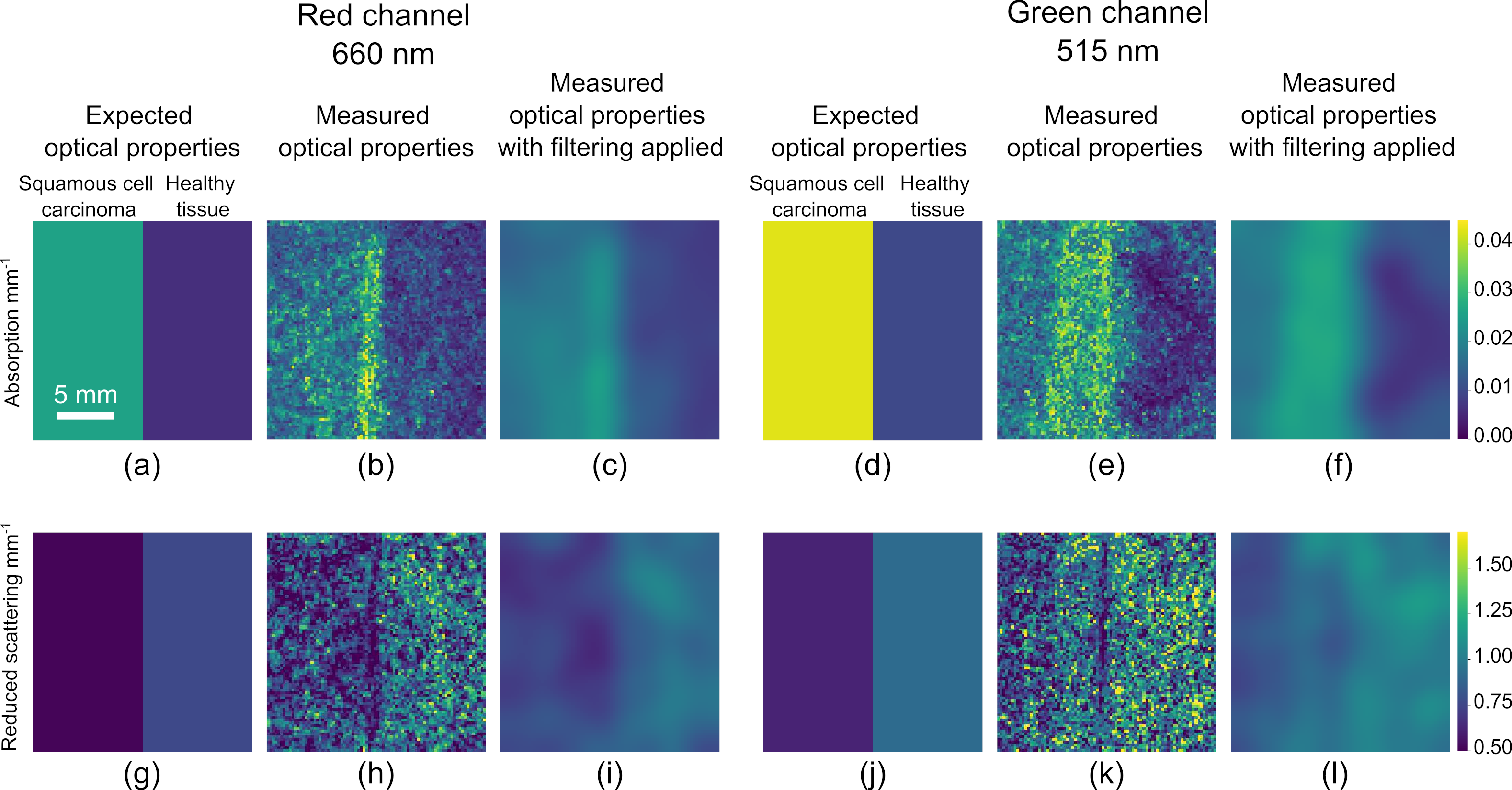}
    \caption{Optical properties measured from dual-wavelength imaging experiment: (a) expected absorption coefficient from red channel (b) measured absorption coefficient from red channel with (c) filtering applied (d) expected absorption coefficient from green channel (e) measured absorption coefficient from green channel with (f) filtering applied. (g) expected reduced scattering coefficient from red channel (h) measured reduced scattering coefficient from red channel with (i)  filtering applied (j) expected reduced scattering coefficient from green channel (k) measured reduced scattering coefficient from green channel with (l) filtering applied.}
    \label{fig:phantom_sidebyside_520and660}
\end{figure}

We then imaged two phantoms with different optical properties placed adjacent to one another, one mimicking the optical properties of squamous cell carcinoma and the other mimicking the optical properties of healthy esophageal tissue. The results are shown in Fig \ref{fig:phantom_sidebyside_520and660} (a-l). The difference in optical properties is visible from both the red and green channels. The optical properties measured from the red and green channel are not expected to be the same as the phantom properties shift with wavelength \cite{Hacker2021}. We expect the phantom optical properties measured from the green channel to be higher than phantom optical properties measured from the red channel. 

\section{Discussion}
\label{sect:Discussion}
We have developed an ultra-miniature SFDI system and shown its capability to quantitatively image differences in optical properties of typical gastrointestinal conditions simulated in tissue-mimicking phantoms, providing enhanced contrast. It is sufficiently small to fit in the instrument channel of a standard colonoscope ($<3$mm). This work could therefore form the basis of new devices suitable for cost-effective endoscopic deployment for screening of gastrointestinal cancers.

This work has limitations that need further investigation before clinical translation. The first limitation is the choice of wavelengths, which in these experiments was $660$ and $515$ nm. By evaluating the absorption coefficient at two wavelengths tissue information such as chromophore concentration can be determined. Oxyhaemoglobin (HbO$_2$) and deoxyhaemoglobin (Hb) are important tissue optical properties because they can detect perfusion, which enables differentiation between malignant and benign tumours\cite{Biswal2011} though wavelengths of $670$ and $850$ nm are more commonly used \cite{Mazhar2010}. Our system has two constraints which make it challenging to extend to the NIR e.g. $850$ nm. First, the micro camera module has an IR filter that blocks light in this range but future versions may remove this. Secondly, the fiber array was designed for $660$ nm, and therefore very lossy when using a $850$ nm laser, with $<$ 1\% efficiency. In future, a fiber array could be designed to operate successfully at both $660$ and $850$ nm: indeed fiber arrays with low-coupling between cores and that operate well into the NIR ($1550$ nm) are routinely used in telecommunications\cite{vanUden2014}.

The second limitation is the need for real time operation for clinical application. Our analysis shows that it typically takes between 0.42s and 2.34s to obtain 3 suitable frames for SFDI (sufficient contrast with 3 shifted phases) while the fiber is being manually perturbed. Though this gives an effective SFDI frame rate of at most 2.4 fps, faster frame-rate cameras could likely improve this: $>100$ fps cameras are widely available. If the system were kept still it would take longer to acquire the 3 desired frames, but in this case artificial phase perturbation could be introduced by means of a phase-shifter or a mode scrambler on one of the fiber arms. However, the phase tracking algorithm is currently relatively slow and runs offline (taking several minutes), so does not allow for real-time operation. This could be addressed by implementing the algorithm on a fast GPU that processes images as they arrive.  Alternatively, images with non-optimal phases could be used for sinusoid fitting instead of waiting for 3 equispaced phases \cite{Greivenkamp1984}. 

The third limitation is image quality, which is somewhat reduced by the non-ideal illumination patterns produced by the fiber array. This causes both noise and some inaccuracies in optical properties.  The noise has been partially addressed by varying the spatial frequency but could be further improved by normalizing for laser power, using AI\cite{Osman2022} for noise-reduction reconstruction or building custom LUTs based on non-ideal projection patterns\cite{Crowley2023Blender}.  The underestimation of absorption at larger absorption levels can be up to 60\%, though this too may be reduced by alternative calibration procedures \cite{Erickson2010}.  This does not appear to significantly impact the quantitative maps of tissue produced, and, more importantly, still permits a suitably high theoretical sensitivity and specificity.  According to the American Society of Gastrointestinal Endoscope (ASGE) guidelines, a new device requires $90$\% sensitivity and $80$\% specificity in order to be used in lieu of biopsy sampling \cite{ASGE2012}.  Post-processing analysis of our data suggest the detection of Barrett's esophagus may fall just short of this, although this could be improved by binning together pixels at the expense of spatial resolution.  However, we show that it is possible to exceed these values when comparing healthy and squamous cell carcinoma models, achieving  $>$99\% sensitivity and 89.6\% specificity. This shows that even in the presence of visually perceptible noise, absorption and scattering when used in combination remain sufficiently robust to provide maps of squamous cell carcinoma.  However, we note that our simulated values of sensitivity and specificity are probably higher than what would be expected in a preclinical study as they do not consider other confounding pathologies such as neoplasias or inflammation.

Further miniaturization of the device could look at the use of metasurfaces for polarizers on the fiber tip\cite{Rodrigues2017}, various fiber tip filters to image different reflected wavelengths\cite{Kim2023}, or patterned surfaces to produce a concentric circle illumination pattern required for wide-field imaging inside tubular lumen.

\section{Conclusion}
\label{Conclusion}
We have shown the capability of an ultra-miniature ($3$ mm diameter) SFDI system to detect quantifiable variances in absorption and reduced scattering coefficients in tissue mimicking phantoms with errors of $15$\% and $6$\% respectively, compared to a conventional bench-top SFDI system. Our device has the capability to project two wavelengths simultaneously, enabling extraction of additional properties such as tissue chromophore information. We fabricated tissue-mimicking phantoms simulating typical gastrointestinal condition of squamous cell carcinoma adjacent to healthy esophageal tissue, where the absorption coefficient of squamous cell carcinoma is much greater than that of healthy tissue and the reduced scattering coefficient is lower. We have shown the capability of our system to image this variation at both one and two wavelengths simultaneously, providing enhanced contrast between the two tissue types. Because of its small size and therefore compatibility with endoscopic deployment, we envisage this system could be used for cost-effective endoscopic screening of gastrointestinal cancers.

\subsection*{Disclosures}
The authors declare no conflict of interest

\subsection*{Acknowledgements}
The authors acknowledge support from a UKRI Future Leaders Fellowship (MR/T041951/1) and an EPSRC Ph.D. Studentship (2268555).

\subsection* {Data, Materials, and Code Availability} 
The data presented in this study are available from the following source: [DOI to be inserted later].


\bibliography{references}   
\bibliographystyle{spiejour}   

\listoffigures
\listoftables

\end{spacing}
\end{document}